\documentclass[a4paper,11pt]{article}
\usepackage{pos}
\newcommand{\ignore}[1]{}

\newcommand{\nn}{\nonumber \\}
\newcommand\be{\begin{equation}}
\newcommand\ee{\end{equation}}
\newcommand\bea{\begin{eqnarray}}
\newcommand\eea{\end{eqnarray}}
\newcommand{\<}{\langle}
\renewcommand{\>}{\rangle}
\newcommand\mR{ \mathbb R}

\newcommand\mS{ \mathbb S}

\graphicspath{{fig/}}
\usepackage{cancel}
\usepackage{lineno}

\title{Ising on $\mathbb{S}^2$ - The Affine Conjecture }
\ShortTitle{The Affine Conjecture }

\author*[a]{Richard C.~Brower}
\author[b]{George T.~Fleming}
\author[c]{Jin-Yun Lin}
\author[a]{Nobuyuki Matsumoto}
\author[a]{Rohan Misra}

\affiliation[a]{Boston University,
  Boston, MA 02215, USA}
\affiliation[b]{Fermi National Accelerator Laboratory,
Batavia, Illinois, 60510, USA}
\affiliation[c]{Carnegie Mellon University, Pittsburgh, PA 15213, USA}

\emailAdd{brower@bu.edu}

\abstract{ We review the recent construction~\cite{brower2024isingmodelmathbbs2} of the 2d Ising model on a triangulated sphere
  $\mS^2$. Surprisingly, this led  to a precise
  map of the lattice  couplings to the target geometry
  in order to reach the   conformal
  field theory (CFT) in the continuum limit.  For the integrable 2d Ising CFT, the map was found analytically~\cite{Brower_2023}.  Here we conjecture how this might be generalized. The discrete geometry
  is implemented  by the  piecewise flat triangulation introduced
  by  Regge in 1960 
   for the Einstein Hilbert action~\cite{Regge1961GeneralRW}.
   Then following our Ising example, we posit the existence of a smooth
    map of lattice couplings in affine parameters consistent with  quantum correlators.
    A sequence of theoretical investigations and numerical simulations are recommended to farther  test  this conjecture. They  begin
   with non-integrable CFT's --  the 2d  $\phi^4$ theory on $\mS^2$; the 3d Ising  model on $\mS^3$ and $\mR \times S^2$; QED3 on $\mathbb R \times \mathbb S^{2} $ as an intermediate  step
   to   4d  non-Abelian lattice gauge  theory on $\mR \times \mS^3$. 
   }

\FullConference{The 41st International Symposium on Lattice Field Theory (LATTICE2024)\\
 28 July - 3 August 2024\\
Liverpool, UK\\}

\begin{document}
\maketitle

\section{Introduction}

The Monte Carlo simulations of the Euclidean path integral in flat space
on hypercubic lattices have proven to be a powerful {\em ab initio}
solutions to non-perturbative field theory as exemplified by lattice QCD. Extending these to lattice field theory on curved manifolds could open up a new area of investigation for non-perturbative quantum field theory. In
1985, Cardy~\cite{Cardy:1985xx} already emphasized the advantage for conformal field theory (CFT) of  radial quantized lattice 
on $\mR \times S^{d-1} $, with the warning of the formidable
challenge of spherical lattices for d >2.  The Quantum Finite Element (QFE) project
has taken on this task, beginning with the Ising CFT or
the universally equivalent  $\phi^4$ theory. In 2d  the stereographic projection to the Riemann sphere, $\mR^2 \rightarrow \mS^2$, is a useful first step to
implementing spherical lattices, 
with  advantage comparison to the exact solution to the $c = 1/2$ minimal model. Development of QFE project is given in a sequence of publications ~\cite{Brower:2016vsl,Brower2018LatticeF,Brower_2021,ayyar2023operator,brower2024isingmodelmathbbs2,Brower_2023}. However in this talk,  technical details
are avoided to provide a heuristic narrative to our conjecture.

\section{Classical Field Limit}
 
We begin with the classical action for  $\phi^4$ theory on a curved Euclidean manifold, 
\be
S_M =   \frac{1}{2} \int_{\cal M} d^dx \sqrt{g}[ g^{\mu\nu}(x) \partial_\mu \phi(x) \partial_\nu \phi(x) +\xi_0 {\bf R} + m^2 \phi^2(x)  + \lambda \phi^4(x)] \; .
\ee
Perturbative renormalization has been extensively studied
for this example~\cite{Luscher:1982wf}. Placing this on a lattice, requires introducing a graph,
assigning the metric ($g_{\mu\nu}(x)$) and field ($\phi(x)$) to sites  $i = 1,..,N$ and directed links  $\<i,j\>$:
\be
S_{FEM}=  \frac{1}{2} \sum_{\<i,j\>}  K_{ij} (\phi_{i} - \phi_{j} )^2
    +  \frac{1}{2}  \sum_{i} \sqrt{g_i} \big[ 
      \mu^2_0  \phi^2_i
     +   \lambda_0   (\phi^2_i -1)^2] \big] \; .
     \label{eq:FEMphi}
\ee
 We
dropped the Ricci scalar ${\bf R}$ which decouples in $2d$ but plays an important role in $3d$ ~\cite{Brower_2021,ayyar2023operator}. On the graph,  all bare parameters are dimensionless. 
The shift in bare mass parameter, $m^2_0 = \mu^2_0 - 2 \lambda_0$, gives
a simple parameterization in the bare coupling $\lambda_0$ 
between the free CFT at $\lambda_0 =0$
to the Ising model at $\lambda_0 = \infty$.  

\subsection{The Regge Simplicial Manifold} 

\begin{figure}[h]
 \centering
 \includegraphics[width = \textwidth]{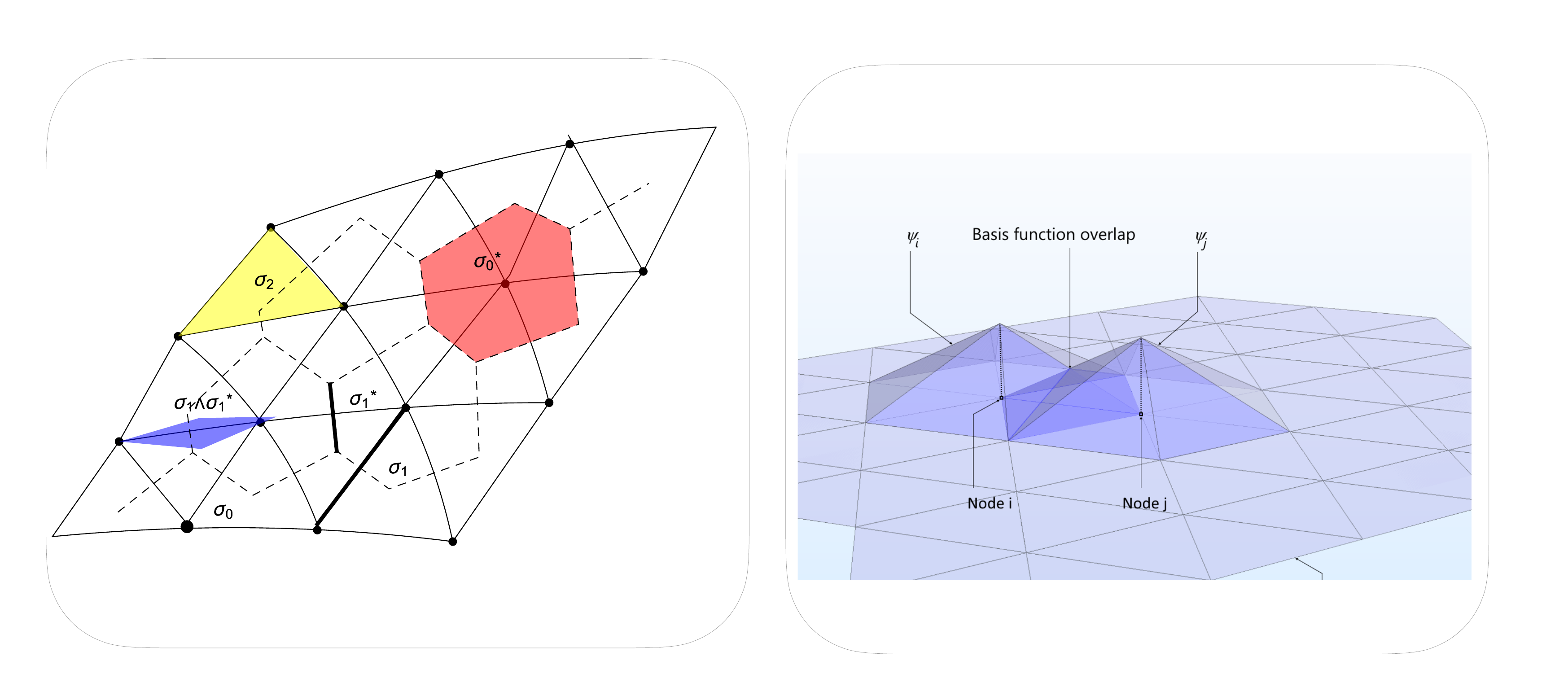}
 \caption{\label{fig:ReggeFEM} On the left is portion of 2d Regge  simplicial
 manifold composed triangular 2 simplices, $\sigma_2(ijk)$,
 joined a boundary edges with length, $|\sigma_1(ij)| = \ell_{ij} $,
 and dual distances, $\sigma^*_1(ij) = \ell^*_{ij}$,  between circumcenter dual sites.
 On right FEM linear basis element for a  scaler field $\phi_i$.}
\end{figure}

At the classical level, a single framework between
geometry and matter is provided by the Einstein Hilbert action,
\be
S = S_{EH} + S_M  = \int dx  \sqrt{g} \big[ \frac{1}{2 \kappa} (  R  -  2 \Lambda) + {\cal L}_M \big ] \; ,
\ee
and its   equation of motion (EOM),
\be
\frac{\delta S}{\delta g^{\mu \nu}(x)} \implies  R_{\mu \nu}(x) - \frac{1}{2} g_{\mu \nu}(x) R(x)  - \Lambda  g_{\mu \nu}(x) =  \kappa T_{\mu\nu}(x) \; .
\ee
To discretize the manifold, we note that Whitney's embedding theorem  states
that any smooth  d-dimensional manifold has an isometric embedding
in $\mR^{2d}$. So Regge approximates the manifold by a set flat planes intersecting the surface, forming d-simplices $\sigma_d$: triangles in 2d,  tetrahedron in 3d etc. This same manifold on the left in Fig.\ref{fig:ReggeFEM}, when used to
discretize  classical equation,  is referred to  as the Finite Element Method (FEM)
illustrated  on the right in Fig~\ref{fig:ReggeFEM}.

\begin{figure}[h]
 \centering
 \includegraphics[width = 0.9\textwidth]{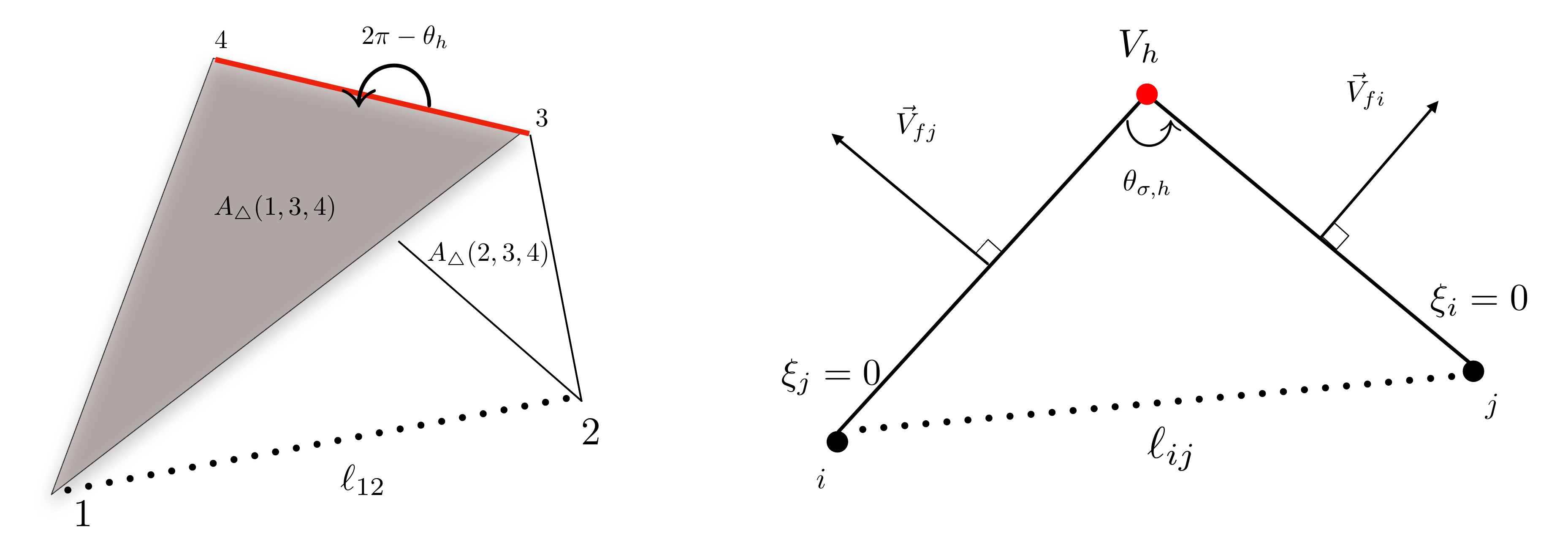}
 \caption{\label{fig:hing} On the left, a tetrahedral simplex
 with a dihedral angle, $\theta_{\sigma,h}$, on the hinge  at the edge $\<3,4\>$ at
 the intersection of two triangles. On the right the hinge for a d-simplex  occupies
  a  $d\!-\!2$ dimensional volume,  $V_h$.  The dihedral angle is defined by a scalar product between vectors  normal to $d\!-\!1$  faces $\vec V_{fi}, \vec V_{fj}$ at $ \xi_i = 0,\xi_j  =0$ respectively.}
\end{figure}

The simplices share boundaries to form a continuous manifold.
The entire  geometry is encoded in the  
flat  interiors of each simplex, $\sigma_d(0,1,..,d)$,  with vertices at $\vec r_i$ in $\mR^d$. The
$d(d+1)/2 $ edge lengths, $\ell_{ij} = |r_i - r_j|$,
determine the affine subspace on each simplex. The  metric
field is replaced by $g_{\mu\nu}(x) \rightarrow \{ \ell_{ij}\}$.
The curvature is isolated to $d\!-\!2$ dimensional delta functions  on hinges weighted by hinge volumes, $V_h $, and the total deficit angle, $\epsilon_h $, from the sum of dihedral angles contributing to the hinge, $h \in \sigma$, as illustrated for 3d in Fig~\ref{fig:hing}. Integrating
the Enstein-Hilbert action yields Regge's calculus action~\cite{Regge1961GeneralRW},
\be
S_{Regge} [\ell_{ij}] = \sum_h V_h \epsilon_h - 2 \Lambda \sum_{\sigma_d} |\sigma_d|\quad \mbox{with} \quad \epsilon_h = 2\pi - \sum_{h \in \sigma} \theta_{\sigma,h} \; ,
\ee
and the equations of motion,
\be
 \frac{\partial S_{Regge}}{\partial \ell_{ij}} = \sum_h\frac{\partial V_h}{\partial \ell_{ij}} \; \epsilon_h + \sum_h\cancelto{ 0}{ V_h  \frac{\partial \epsilon_h}{\partial \ell_{ij}}}   - 2 \Lambda \sum_{\sigma_d \supset \ell_{ij}} \frac{\partial |\sigma_d|}{\partial \ell_{ij}} = 0 \; .
\ee
The second term  in the middle is exactly zero due the Schl\"{a}fli identity~\cite{Schlafli}.
This is a mathematically elegant FEM  discretization~\cite{barrett2019tullioreggeslegacyregge}
with discrete curvature tensors, Bianchi identities~\cite{Hamber_2004,ariwahjoedi201721reggecalculusdiscrete,Ariwahjoedi:2018qbr} etc. With
positive cosmological constant, Ricci flow~\cite{Miller_2014} converges 
to spherical manifolds. Apparently
the  Regge 3d manifold also helped to illuminate the proof of the Poincar\'e conjecture~\cite{perelman2003ricciflowsurgerythreemanifolds,morgan2007ricciflowpoincareconjecture} that there is unique manifold homeomorphic to $\mS^3$.

Turning to the scalar field theory in 2d,  the
 weights  in Eq.~\ref{eq:FEMphi} on edge $\<i,j\>$ are given by
\be
K_{ij} = \frac{\ell^*_{ij}}{\ell_{ij}}
\label{eq:free_phi}
\ee
using piecewise linear FEM, where  $\ell^*_{ij}$ are distance on the circumcenter dual lattice.

For a simple example  of the intimate relation between
the Regge's  geometry  and linear finite elements,
we now sketch the computation of kinetic term (\ref{eq:free_phi})  for 
$d \ge 2$. The central trick is to introduce  barycentric co-ordinates, $\xi_i$,  both for  the interior of a d-simplex, $\sigma_d(0,1,...,d)$, with $\xi_k \le 0$ and $\xi_0 + \xi_1 + \cdots + \xi_d= 1$
and for the linear interpolation field values: $\phi_i = \phi(\vec r_i)$ at the sites $\vec r_i$:
\be
\vec x  = \xi^0 \vec r_0 + \xi^1 \vec r_1 + \cdots +\xi^d\vec r_d 
\quad,\quad \phi(x) = \xi^0 \phi_0  + \xi^1 \phi_1 + \cdots + \xi^d \phi_d
\ee
Changing variables in the integral  we compute the
simplex contribution to the  edges weights,
\be
\int_{\sigma_d} d^dx \sqrt{g} g^{\mu \nu}\partial_\mu \phi(x) \partial_\mu \phi(x) \quad  \implies \quad
K^{\sigma_d}_{ij} = - |\sigma_d| \vec\nabla \xi^i \cdot \vec \nabla \xi^j  =
\frac{\vec V_{f_i} \cdot \vec V_{f_j}}{d^2 |\sigma_d|} = \frac{\partial V_d}{\partial \ell^2_{ij}} \; .
\ee
The FEM kinetic term
is geometrized as the derivative of the volume $V_d = |\sigma_d|$
with respect to the edge conjugate to the hinge! A slight generalization
gives the  energy  momentum tensor as well. The trace matches
the cosmological term. This is just a taste of the
remarkable identities.  
For the free CFT ($\lambda_0 = 0$) Regge plus FEM
is a complete solution -- both the geometry and the field extrapolate to exact
continuum theory in the limit $\ell_{ij} < a$ as $a \rightarrow 0$.
{\bf Now our task is to find a way to transfer this simplicial geometry 
to lattice  quantum field theory correlators.}

\subsection{Affine Space}

The affine geometry plays  a key role in our generalization.
In  $\mR^d$  the  affine transformation is  general linear 
map,
\be
x^\mu = A_{\mu,i}\xi^i + b^\mu \quad \mbox{or} \quad \vec x = \vec e_i \xi_i + \vec b
\ee
extending the $d(d+1)/2$ Poincare generators by a factor of 2  to
a total of $d(d+1)$ generators.
The additional  parameters represent a constant affine metric,
\be
ds^2 = d\vec x \cdot  d\vec x = (A^T A)_{ij} d\xi^i d\xi^j = g_{ij} d\xi^i d\xi^j  \quad \implies \quad  g_{ij} = \vec e_i \cdot \vec e_j \;.
\ee
Each simplex is affine equivalent to a standard equilateral simplex with unit edge lengths. For 2d the $d(d+1)/2 = 3$ parameters match  the triangle edges lengths, $\ell_{12}, \ell_{23}, \ell_{31}$ -- representing scale (similarity)  plus shape.
In 3d the  tetrahedron has a 6 affine edge lengths, the 4-plex 10 edge lengths, etc. Regge constructs the simplicial geometry by gluing
the affine subspaces into a piecewise linear manifold. 
We also anticipate that a global affine transformation for
lattice field correlators at  the critical point takes circles (spheres)  to ellipses (ellipsoids). For 2d, this was proven 
\be
\<\phi(x,y)\phi(0,0) \> = \frac{1}{(x^2 + y^2)^{\Delta_\phi}} \rightarrow \frac{1}{(a x^2 + b y^2 + c xy)^{\Delta_\phi}} \; ,
\ee
for the general triangular Ising model (\cite{Brower_2023}).

\section{Ising model on a 2-sphere}
\label{sec:IsingSphere}

In Ref~\cite{Brower_2023}  on the {\em Ising Model on the Affine Plane }, we
gave an analytical solution to  the general triangular Ising model with 3 couplings  $K_1,K_2,K_3$. 
\begin{figure}[h]
 \centering
 \includegraphics[width = 0.7\textwidth]{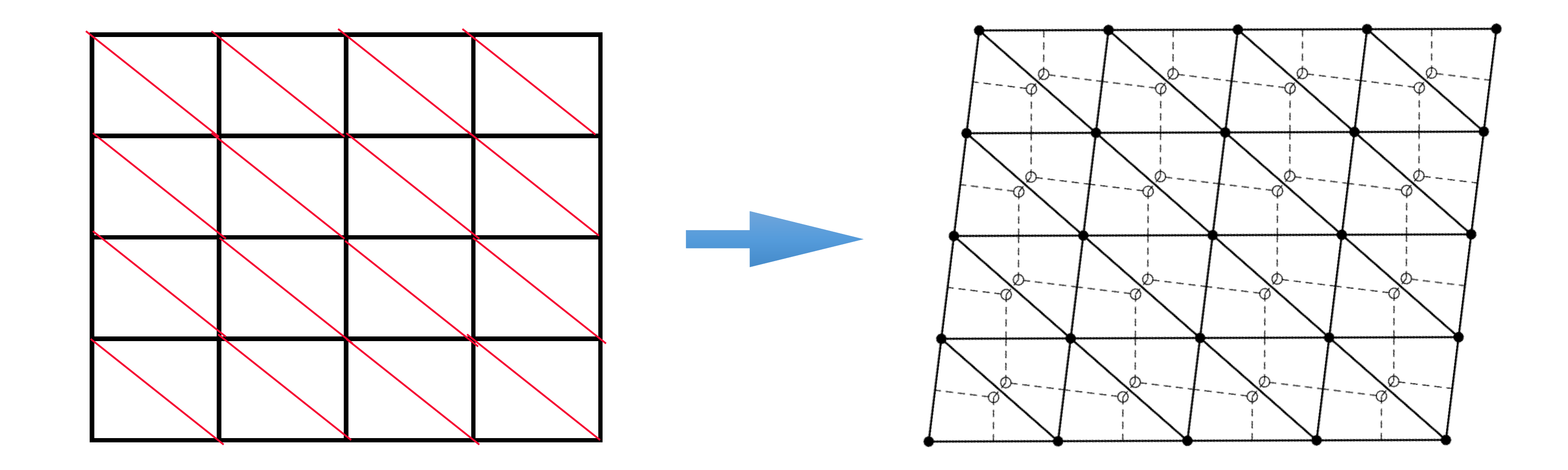}
 \caption{\label{fig:AffineIsing} On the square lattice
 Ising  model, $S_\square =  -  K (s_i s_{i+\hat 1} +s_i s_{i + \hat 2})$, on the left  is generalized  to a regular triangular graph, $S_\triangle =  -   K_1 s_i s_{i + \hat 1} - K_2 s_i s_{i + \hat 2} -  K_3 s_i s_{i + \hat 3}$ with  3 couplings. }
\end{figure}
The solution follows from the star-triangle relation, the Kramers-Wannier
map to hexagonal lattice and the use free Wilson-Majorana fermions
as describe by Wolff in Ref.~\cite{Wolff2020IsingMA}.  To restore
spherical symmetry in $\mR^2$ the metric map along
3  oblique co-ordinate is
\be
\sinh(2 K_1) = \frac{\ell^*_1}{\ell_1} \quad, \quad \sinh(2 K_2) = \frac{\ell^*_2}{\ell_2} \quad, \quad \sinh(2 K_3) = \frac{\ell^*_3}{\ell_3} \;. 
\ee
The 3  scale-invariant equations implying the constraint to 
the 2d critical surface,
\be
p_1 p_2 + p_2 p_3 + p_3 p_1 = 1 
\quad \mbox{with} \quad  p_i = e^{-2 K_i} \; .
\ee

\begin{figure}[h]
 \centering
 \includegraphics[width = \textwidth]{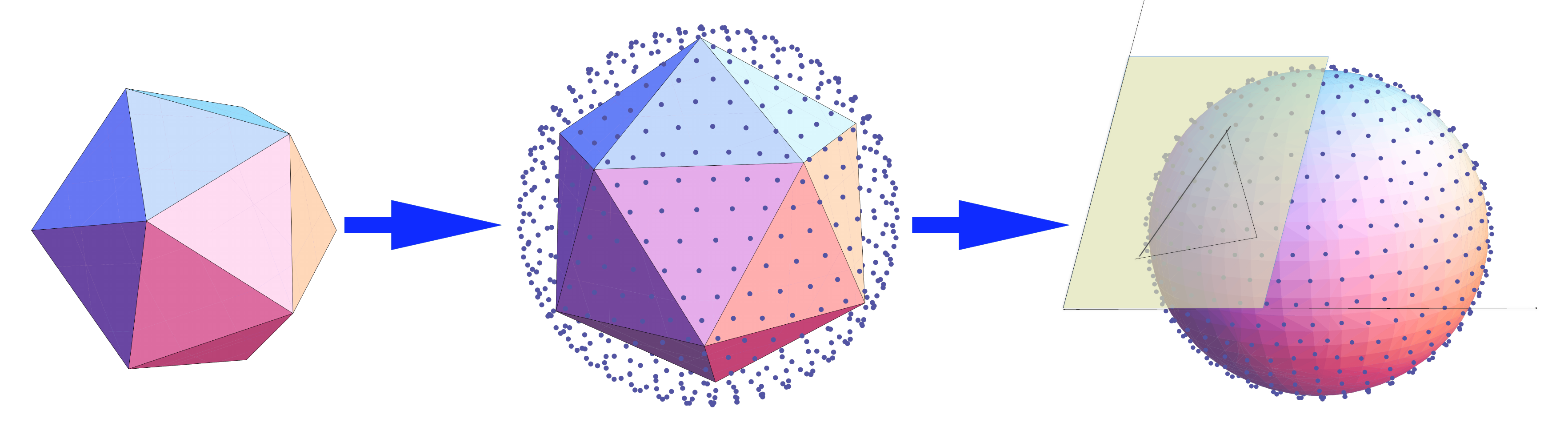}
 \caption{\label{fig:icos_refine}Steps in the basic discretization of $\mS^2$ using an icosahedral base refining-- shown here for a refinement of L = 3 into 
    $L^2$ triangle on each of the 20 icosahedral faces, subsequently projected onto the sphere. The Euler condition, $N - E + F = 2$, is
    satisfied for $N = 2 + 10 L^2$  (sites) ,  $E = 30 L^2$ (edges)  and $F = 20 L^2$  (faces). }
\end{figure}

Moving to a smooth triangulation of 
the 2-sphere 
introducing equilateral triangles on each of the 20 faces of the Icosahedron
projecting to radially unit  3-vectors $\vec r_i$ in $\mR^3$ as illustrated
in Fig.\ref{fig:icos_refine}. The result is
smooth but non-uniform triangulation of the sphere that in the
continuum limit approaches an affine map to each tangent plane.
We introduce  a nearest neighbor Ising  model 
on the triangular lattice,
\be
S_{\mS^2} = -\sum_{\<i,j\>} K_{ij} s_i s_j \quad \mbox{with} \quad  \sinh(2 K_{ij}) = \frac{\ell^*_{ij}}{\ell_{ij}} \; ,
\ee
with coupling constraint form  as you approach
the continuum locally on each tangent plane as a function of the edge lengths, $\ell_{ij}= |\vec r_i - \vec r_j|$,
and its circumcenter dual lengths, $\ell^*_{ij}$, perpendicular to the
edge $\<i,j\>$.   The Icosahedron has 120 element Icosahedron subgroup $I_h$ of $O(3)$, which is respected by our  {\em naive}  icosahedral 
projecting and our subsequent equal area  refinement (\ref{eq:EqualArea}).

\section{Numerical Test on $\mS^2$}

As a comparison we recall earlier we introduce  counter terms, for the $\lambda_0 \phi^4$ theory to remove the non-uniform UV cut-off on the triangulated sphere. 
The  kinetic  term (\ref{eq:free_phi}) is unchanged for the FEM form,$K_{ij} = \ell^*_{ij}/\ell_{ij}$, which is exact
for the free theory ($\lambda_0 = 0$). The FEM form
is modified by  locally mass shift by $\delta m^2_i =  \lambda_0 \log(a^2/a^2_i) $ by numerically
computing the one loop the UV divergent diagram  on the triangulated sphere.
Even at $\lambda_0 = 1$ Monte Carlos simulations  gave  remarkably accurate results on both are $\mS^2$ and with similar 3d counter terms on $\mR \times \mS^2$. Further analysis demonstrated that in the continuum
limit the bare coupling must be scaled
to zero holding the renormalized coupling $\lambda_R \sim \lambda_0/a^2 $ fixed. 

\begin{figure}[h]
   \centering
   \includegraphics[width =0.49\textwidth]{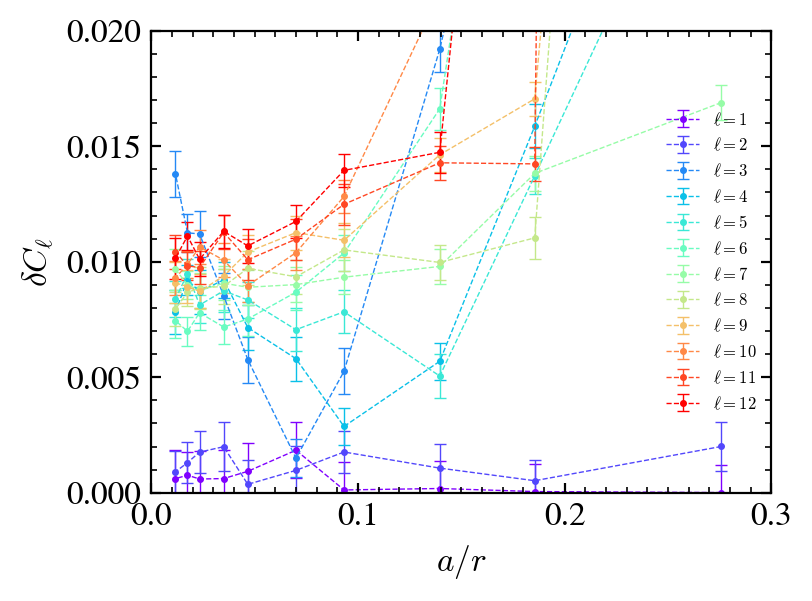}
    \includegraphics[width =0.49\textwidth]{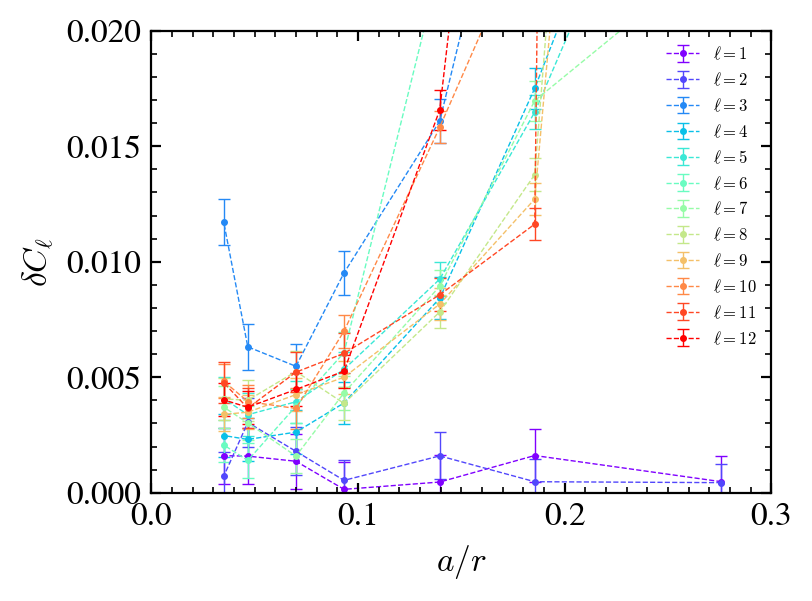}
    \caption{Breaking of rotational symmetry using the basic icosahedral discretization of $S^2$. On the left for the Ising module vs on the right for the critical $\phi^4$ theory with local perturbative counter-terms }
    \label{fig:basic_break}
  \end{figure}

Where as for the Ising model we used coupling 
constants, $\sinh(2 K_{ij})  = \ell^*_{ij}/\ell_{ij}$,
consistent with the $\lambda_0 =\infty$
The  result is a UV critical theory with only
nearest neighbor coupling and NO counter terms. Testing restoration
of spherical symmetry for the two point function is
given in Fig.\ref{fig:basic_break}. 

We note that  our critical Ising map on the left in Fig.~\ref{fig:basic_break} is if anything {\bf worse} than for $\phi^4$ theory with perturbative counter term? {\bf What is wrong?}  
Typically the FEM for the Regge manifold is  very forgiving. 
Any  smooth triangulation of the sphere approaches
the continuum manifold with a differentiable metric. Indeed in Ref.~\cite{Feinberg:1984he} it
was proven that the  deficit $\epsilon_h$ on each  hinge is proportional dual areas $A^*_h$ at the hinge: $\epsilon_h \sim A^*_h + O(a^2)$. By  the Gauss-Bonnet theorem  this guarantees exact convergence to our $\mS^2$ geometry.
However to reduce  $O(a^2)$ cut-off effects, it 
is plausible to also force the curvature singularities,  $\epsilon_h$, to have equal strength.

\begin{figure}[h]
   \centering
   \includegraphics[width = 1.1\textwidth]{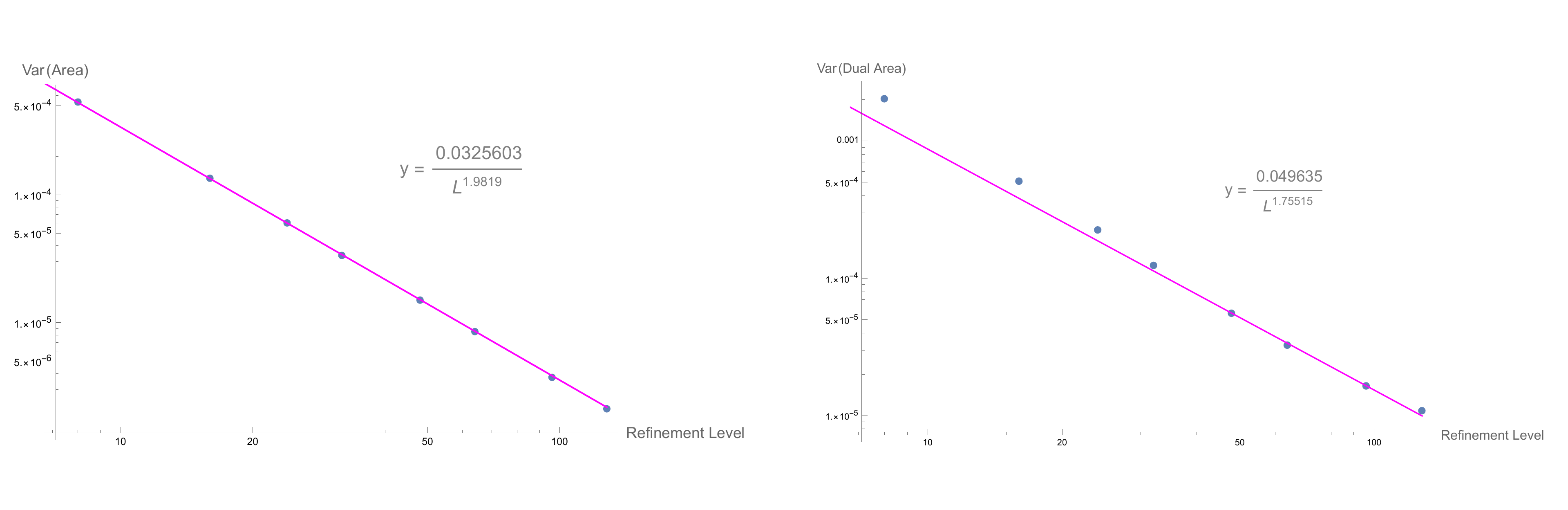}
    \caption{On the left the RMS value of triangle area as
    the refinement increase and on the right the consequences for 
    the Dual area.}
    \label{fig:SmoothArea}
  \end{figure}

\begin{figure}[ht]
   \centering
   \includegraphics[width =0.5\textwidth]{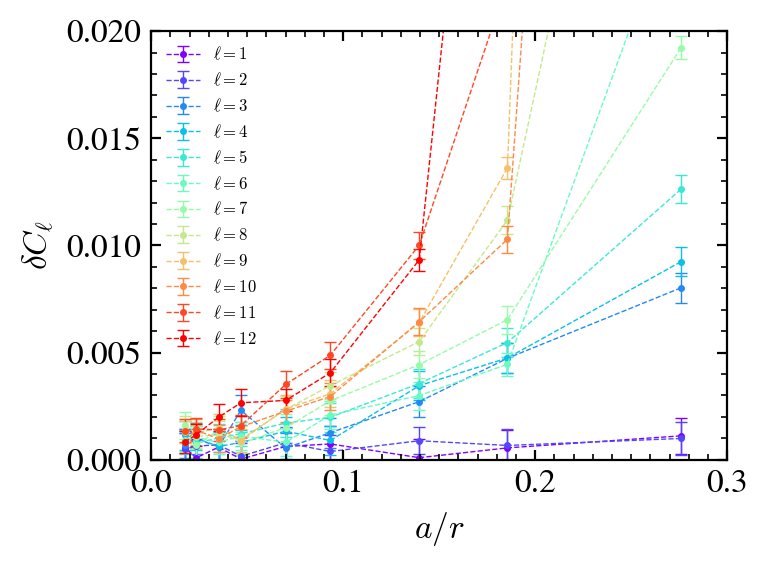}
    \includegraphics[width =0.49\textwidth]{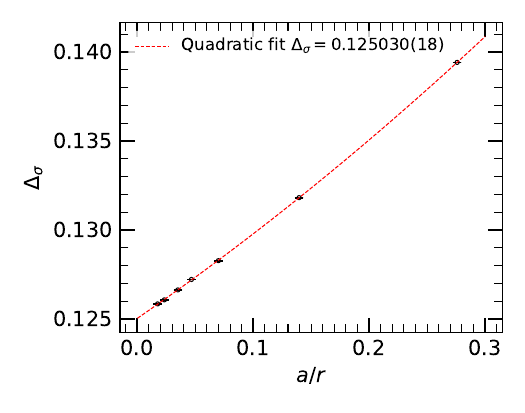}
    \caption{\label{fig:CFT_test} On the left reduced spherical symmetry break 
    approaching the continuum with  equal  Area co-ordinate smoothing.  On the right the Ising $Z_2$ scalar dimension approaches  the exact value $\Delta_\sigma = 1/4$ to $O(10^{-4})$.}
    \label{fig:SoothRotation}
  \end{figure}

To accomplish this, we vary the edge length $\ell_{ij}$ to  minimize the
squared area of  the triangle $A_\triangle(i^*)$
\be
E_\triangle[ \ell_{ij}] = \frac{1}{N} Min[ \sum^{F}_{i^* = 1}  A^2_{\triangle}(i^*)]
\label{eq:EqualArea}
\ee
labeled by the dual sites $i^* = 1,2,\cdots , F $ (or F faces) by moving the unit vector   $\hat r_i$ for position the site. Since the co-ordinates, $\hat r_i$,  have 2N degrees of freedom with $F = 4+ 2N^2$, after removing 3 rotations and one scale, this is 
1-1 constrained system for a non-linear~\footnote{The area squared using Heron's formula, $16 A^2(a,b,c) = (a+b+c)(-a+b+c)(a-b+c)(a+b-c)$, with 3  edge lengths $\ell^2_{ij} = 2 - 2 \hat r_i \cdot \hat r_j $ labeled by $a,b,c$ is an 8th order polynomial in the spins.} ferromagnetic  Heisenberg  spin system $\vec S_i = \hat r_i $. Most likely, there is a unique ground state  in the continuum as illustrated on the left in Fig.~\ref{fig:SmoothArea}. Note if instead we had chosen to minimize the dual $N = 10L^2$
areas,  this system would be ill-determine. Fortunately area minimization accomplishes this smoothing of curvature density as a side effect
as illustrated on the right in Fig.~\ref{fig:SmoothArea}.
 With area smoothing we now see that spherical symmetry is improve substantially in Fig.\ref{fig:CFT_test} on the left
and  on the right, the fits to the two point function agrees with the exact value with error of $10^{-4}$ with modest simulations.

With the exact Icosahedral group preserved in our
smoothing we are now improving the tests rotational symmetry
by imposing this on the Monte Carlo sample
\bea
C_{l_1 m_1; l_2m_2} &=& \sum_{i,j} \sqrt{g_i} Y^*_{l_1m_1}(\hat r_i)  \frac{1}{|g|} \sum_{ g \in Ih}  \langle s(g \hat r_i) s(g \hat r_j)  \rangle \sqrt{g_j}Y_{l_2,m_2}(\hat r_j) \nn
 &\rightarrow&  c_{l} P_{l_1}(\hat r_i \cdot \hat r_j)  \delta_{l_1,l_2}  \delta_{m_1,m_2}+O(a^2) 
\eea
a average over the Icosahedral group.  This will remove exactly the sampling errors in the IR representations
for the $l = 0,1,2$ levels  visible in
Fig.~\ref{fig:basic_break} and Fig.\ref{fig:SoothRotation}. With higher statistic 
on lattice with smaller lattices spacing, we believe tests to a $O(10^{-5})$ tolerance are feasible.

\section{Next steps}
Beyond the 2d Ising model, we are extending
our methods to non-integral field theories.
First for the 2d $\phi^4$ on $\mS^2$ lattices  and  
next for 3d Ising and $\phi^4$ theory on
$\mR\times \mS^2$ and $\mS^3$. Remarkable 
$\mS^3$ has an  analogue of the  Icosahedral
Platonic. The  convex regular 4-polytope 600 cell with Schl\"{a}fli~\cite{Schlafli} symbol $\{3,3,5\}$ or  600 cell  with $120^2 = 14400$ elements by embedding two copies of Icosahedral group in $SO(4) = SU(2)_L \times SU(2)_R/Z_2$. 
To introduce gauge and fermion
beyond the FEM form~\cite{Brower:2016vsl}, we are implementing
$QED3$ with even number of flavors. The question of whether 2 flavor is conformal or not is not yet settled~\cite{Karthik_2016}. This is nice 3d analogue
to the extension to 4d gauge multi-flavor near conformal BSM
theories~\cite{Brower_2016}.

In quantum field theory, we know
that the Energy Momentum operator is the stress in response to changes in metric stain for correlators:
\be
\frac{\delta \langle \phi(x_1) \phi(x_2) \cdots \phi(x_n) \rangle_g }{ \delta g^{\mu \nu}(x)}  = \langle \phi(x_1) \phi(x_2) \cdots \phi(x_n) T_{\mu \nu}(x) \rangle_g  
\ee
The fundament problem in Euclidean space is to lift the
stress strain analysis~\cite{landau1986theory} of  classical FEM to a
quantum critical point. We are considering shifted
boundary methods\cite{Giusti_2011}
and renormalized  Wilson flow operators~\cite{PhysRevD.91.074513}
to design better  algorithmic methods and provide
a stronger theoretical understanding of the affine map to
lattice quantum fields on curved manifolds.

\begin{acknowledgments}
This work was supported by the U.S. Department of Energy (DOE) under Award No. DE-SC0019139,  Award No. DE-SC0015845  and at
FNAL  under Contract
No. 89243024CSC000002.
\end{acknowledgments}

\bibliographystyle{JHEP}
\bibliography{bib/IsingOnS2}

\end{document}